\documentclass[10pt] {article}
\usepackage {enumitem}
\usepackage[T1]{fontenc}
\usepackage[utf8]{inputenc}
\usepackage[paperheight=29.7cm,paperwidth=21.0cm,left=3.17cm,right=3.17cm,top=2.54cm,bottom=2.54cm]{geometry}
\usepackage{indentfirst} 
\usepackage{cite}
\usepackage{hyperref}
\usepackage{amsmath}
\usepackage{titlesec}
\usepackage{amssymb}
\usepackage{float}
\usepackage{adjustbox}
\usepackage{multicol}
\usepackage{multirow}

\begin{document}

\begin{center}
	{\LARGE\textbf{Uncertainty-driven and Adversarial Calibration Learning for Epicardial Adipose Tissue Segmentation}}
\end{center}
	
\vspace{1\baselineskip}
\begin{center}
Kai Zhao, Zhiming Liu, Jiaqi Liu, Jingbiao Zhou, Bihong Liao, Huifang Tang,
\end{center}

\begin{center}
Qiuyu Wang, Chunquan Li\textsuperscript{$\ast$}
\end{center}

\vspace{2\baselineskip}
{\Large \textbf{Abstract}}
Epicardial adipose tissue (EAT) is a type of visceral fat that can secrete large amounts of adipokines to affect the myocardium and coronary arteries. EAT volume and density can be used as independent risk markers measurement of volume by noninvasive magnetic resonance images is the best method of assessing EAT. However, segmenting EAT is challenging due to the low contrast between EAT and pericardial effusion and the presence of motion artifacts. we propose a novel feature latent space multilevel supervision network (SPDNet) with uncertainty-driven and adversarial calibration learning to enhance segmentation for more accurate EAT volume estimation. The network first addresses the blurring of EAT edges due to the medical images in the open medical environments with low quality or out-of-distribution by modeling the uncertainty as a Gaussian distribution in the feature latent space, which using its Bayesian estimation as a regularization constraint to optimize SwinUNETR. Second, an adversarial training strategy is introduced to calibrate the segmentation feature map and consider the multi-scale feature differences between the uncertainty-guided predictive segmentation and the ground truth segmentation, synthesizing the multi-scale adversarial loss directly improves the ability to discriminate the similarity between organizations. Experiments on both the cardiac public MRI dataset (ACDC) and the real-world clinical cohort EAT dataset show that the proposed network outperforms mainstream models, validating that uncertainty-driven and adversarial calibration learning can be used to provide additional information for modeling multi-scale ambiguities.

\vspace{1\baselineskip}

\textbf{KeyWord: Epicardial adipose tissue segmentation, Feature latent space, Uncertainty-driven, Adversarial calibration, Bayesian estimation}

\indent\setlength{\parindent}{1em}
\section*{Introduction}
Epicardial adipose tissue (EAT) resides within both the atrioventricular and interventricular grooves, situated proximal to the coronary arteries and myocardium, enveloping a considerable portion of the cardiac vasculature. As the volume of EAT increases, adipose deposits commence extending towards the ventricular free wall, potentially enshrouding the entirety of the cardiac organ. EAT functions as a metabolically active endocrine and paracrine organ, releasing significant quantities of both anti-inflammatory and pro-inflammatory adipokines. These bioactive molecules exert influence over the dynamic equilibrium of energy-glycolipid metabolism, as well as cardiac structure and function\cite{r1,r2}. EAT is closely associated with the development of coronary heart disease, thus it is considered to be a biomarker for a class of risk factors. With advances in imaging technology, it has been possible in recent years to determine the natural evolution of its volume and density in at-risk populations, suggesting that EAT might undergo substantial alterations in plaque calcification well in advance of its maturation. More findings propose that EAT serves as an independent parameter that is expected to be a surrogate biomarker for cardiovascular risk estimation\cite{r3,r4}.

Nevertheless, the quantitative imaging assessment of EAT with intricate topology is a challenging task. Since EATs have different physiologic and pathologic properties at different locations, this leads to two major problems for the imaging physician in accurately quantifying their statistical differences: Primarily, the heart is a continually pulsating and contracting muscular organ that plays a pivotal role in pumping blood to sustain proper bodily function. This constant motion produces artifacts, leading to the indistinct delineation of EAT edges from the surrounding tissues, as illustrated in Fig. 1a. Secondly, there exists a challenge in distinguishing EAT from pericardial effusion, as both may manifest very similar signal intensities in MRI\cite{r5}, as depicted in Fig. 1b. Consequently, this makes the difference in their grayscale values small and increases increasing the difficulty of visual differentiation. In addition, the morphology and distribution of EAT may exhibit variations based on individual characteristics, further increasing the difficulty of accurately identifying tissue edges. The above factors lead to inefficient and cognitively subjective manual labeling assessments. Thus, to reduce the burden of quantitative assessment of EAT using image processing techniques for automated operations can provide objective and rapid quantitative assessment, which greatly improves the efficiency of imaging physicians.

\begin{figure}
    \centering
    \includegraphics[width=1\linewidth]{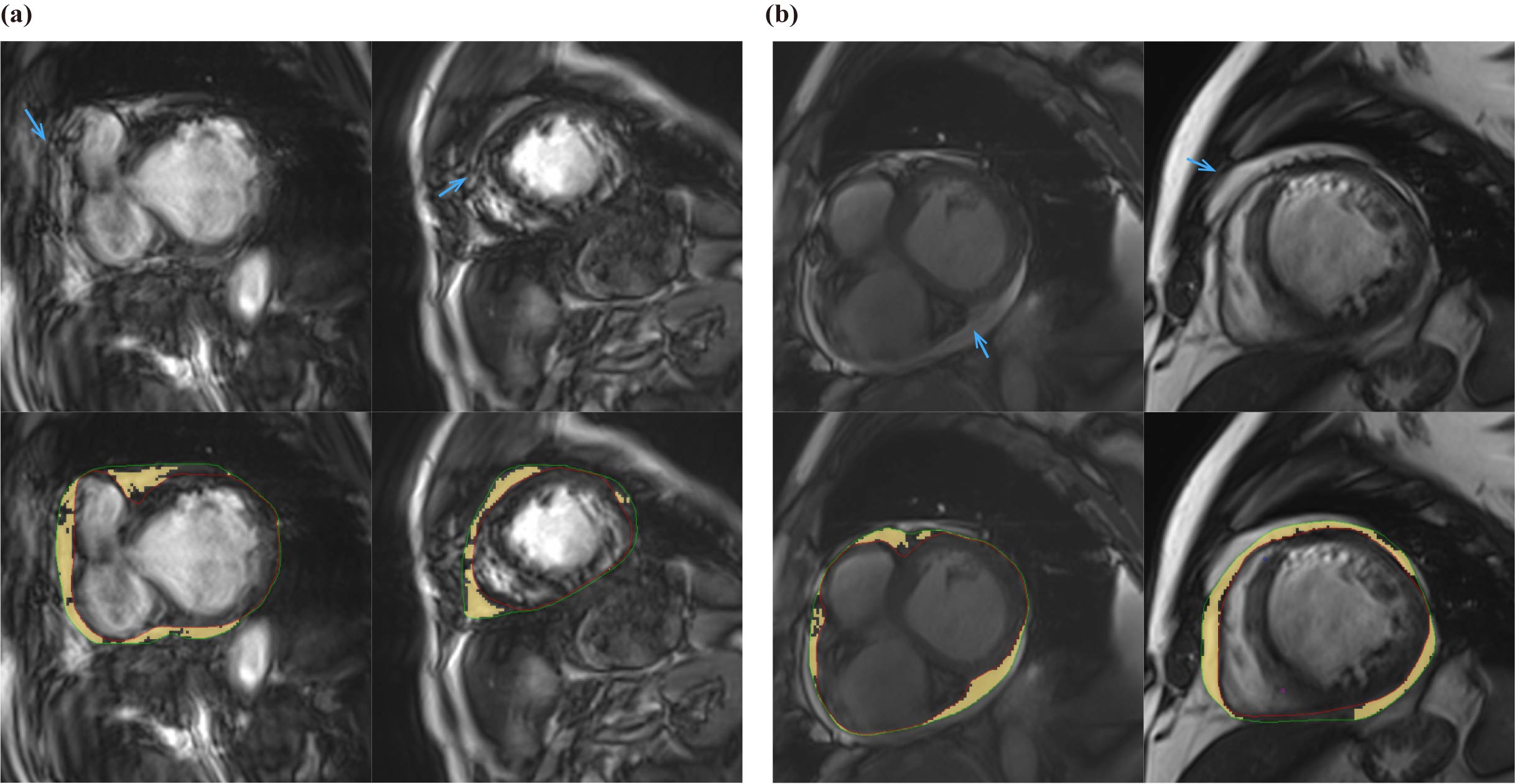}
    {\footnotesize \textbf{Fig1.}}
    {\footnotesize The first row shows four slices of the heart at different times during the MRI scan, and the second row of images marks the region of epicardial fat in yellow. The first two samples show the situation when motion blur is present, and the last two samples show the situation when both pericardial effusion and epicardial fat are present. We have pointed out some of the corresponding regions with blue arrows.}
    \label{fig:enter-label}
\end{figure}

In the field of medical image segmentation, most of the current mainstream semantic segmentation methods use Vision Transformer\cite{r6} or U-Net\cite{r7} architectures, focusing on the learning of deterministic features, e.g.\cite{r8,r9,r10}. Regrettably, these methodologies overlook the influence of data or model uncertainty issues on segmentation outcomes, encompassing factors like motion artifacts and cognitive ambiguities. So, there may be certain model performance bottlenecks for segmentation problems with such uncertainties. Although there have been some approaches that consider the reliability of segmentation results and introduce uncertainty estimation, where uncertainty is directly or indirectly represented as network parameter distributions or feature latent space distributions in the model, these models are based on learning strategies for a parameter or feature distributions do not consider calibrating the modeled uncertainty\cite{r11,r12,r13}, and lack the capability to further utilize uncertainty to guide the segmentation performance improvement. In this study, an uncertainty quantification and calibration framework for feature latent space distribution representation is directly constructed based on the above methods: First, the segmentation model adopts SwinUNETR as its backbone network. Subsequently, a Probabilistic Net based on Bayesian estimation is constructed to drive the uncertainty representation of the model. Finally, a Discriminator Net is constructed to perform adversarial calibration of the segmentation feature maps and to enhance the network's utilization of multi-scale features. Our objective is to investigate an efficient and automated method for EAT segmentation.

The main contributions of this study are as follows:(1) The Probabilistic Net have engineered adeptly models data uncertainty by employing a Gaussian distribution within the feature latent space. Feature vectors sampled from this distribution are incorporated into the segmentation feature maps layer by layer and by using quantized uncertainty constraints to suppress segmentation performance degradation due to motion artifacts. (2) The Adversarial training strategy was used to calibrate the segmentation feature maps, and a multi-scale loss optimization model of Discriminator Net was introduced into the objective function to enhance the model's ability to capture the feature differences at each level, to better distinguish similar tissues with low contrast. (3) The model proposed in this study was validated on an open data set for real-world applications, and the efficacy of uncertainty-driven and adversarial calibration methodologies underwent evaluation in publicly available datasets and self-constructed clinical cohort segmentation tasks. The results show that the proposed method achieves significant improvements in EAT segmentation, which provides strong support for early diagnosis and treatment of coronary artery disease.

\section*{Related work}
\textbf{Probability Model for Images Segmentation:} Image segmentation tasks aim to assign a category label to each pixel in an image. However, in the real-world implementation, a large amount of information is inherently uncertain, leading to potential inconsistencies in the distribution of training and test data. Deterministic segmentation methods encounter difficulties in resolving such uncertainties. Rahman introduced approximate Bayesian inference to model this inconsistency in their study, demonstrating its effectiveness, particularly with smaller datasets\cite{r14}. Kohl proposed the Probabilistic U-Net\cite{r11}, which integrates a conditional variational autoencoder (CVAE)\cite{r15} with a U-Net. This innovative approach addresses the inconsistency issue by leveraging the stochastic nature of the conditional variational autoencoder to generate multiple segmentation hypotheses. However, this method's introduction of uncertainty information solely in the topmost layer of the U-Net encoder leads to limitations in explaining features at different scales. To address this, PHi-Seg improve the model structure, acquiring a hierarchical latent space by the feature of the image at various scales, and infusing feature sampled from this space into each layer of the encoder\cite{r16}. This adaptation enables the latent space to adapt to changing image dimensions, resulting in a improvement in the model's performance, particularly in capturing intricate details. These methods mentioned above have significant advantages when faced with uncertainty in segmentation problems.

\textbf{Adversarial Training Strategies:} The adversarial training strategy was initially introduced in Generative Adversarial Network (GAN)\cite{r17} by Goodfellow et al. This strategy enhances the generalization ability of the model by optimizing the adversarial game between the discriminator and the generator to generate more realistic samples. In the field of image segmentation, models employing this strategy typically consist of two network modules: a segmentation network (Generator) dedicated to image segmentation and a discriminative network (Discriminator) tasked with discerning between real and generated labels. For instance, Xu et al. introduced a semi-supervised adversarial approach, leveraging an adversarial training strategy to enhance segmentation prediction accuracy by enabling the network to learn the data distribution of labels\cite{r18}. Hung et al. combined cross-entropy loss with adversarial loss, resulting in a substantial improvement in the semantic segmentation accuracy of the model\cite{r19}. The findings presented in\cite{r20} demonstrate that the adversarial training strategy proves effective in mitigating labeling noise in real-world data and performs admirably in scenarios with limited dataset sizes. Furthermore, in\cite{r21} noteworthy is the incorporation of multi-scale loss functions in discriminator networks, seamlessly integrated into segmentation networks through adversarial training. This enables them to grasp both global and local features, leading to heightened model accuracy. These research outcomes unequivocally underscore the validity of adversarial training strategies within the realm of image segmentation, offering substantial validation for our research endeavors.

\section*{Materials and methods}
Fig. 2 shows the model structure of the methodology of this study: (1) the SwinUNETR-based backbone network, denoted as Segmentor Net, (2) the uncertainty-driven Probabilistic Net, and (3) Discriminator Net for calibrating uncertainty representations of feature maps and computing multi-scale loss.

The proposed methods are slated for validation using two distinct datasets: the public ACDC MICCAI 2017 dataset\cite{r22} and a real-world clinical cohort dataset on EAT. Detailed elucidation of every network and their optimization algorithms that exploit the uncertainty representation of the feature latent space and adversarial training calibration are described in detail below.

\begin{figure}
    \centering
    \includegraphics[width=1\linewidth]{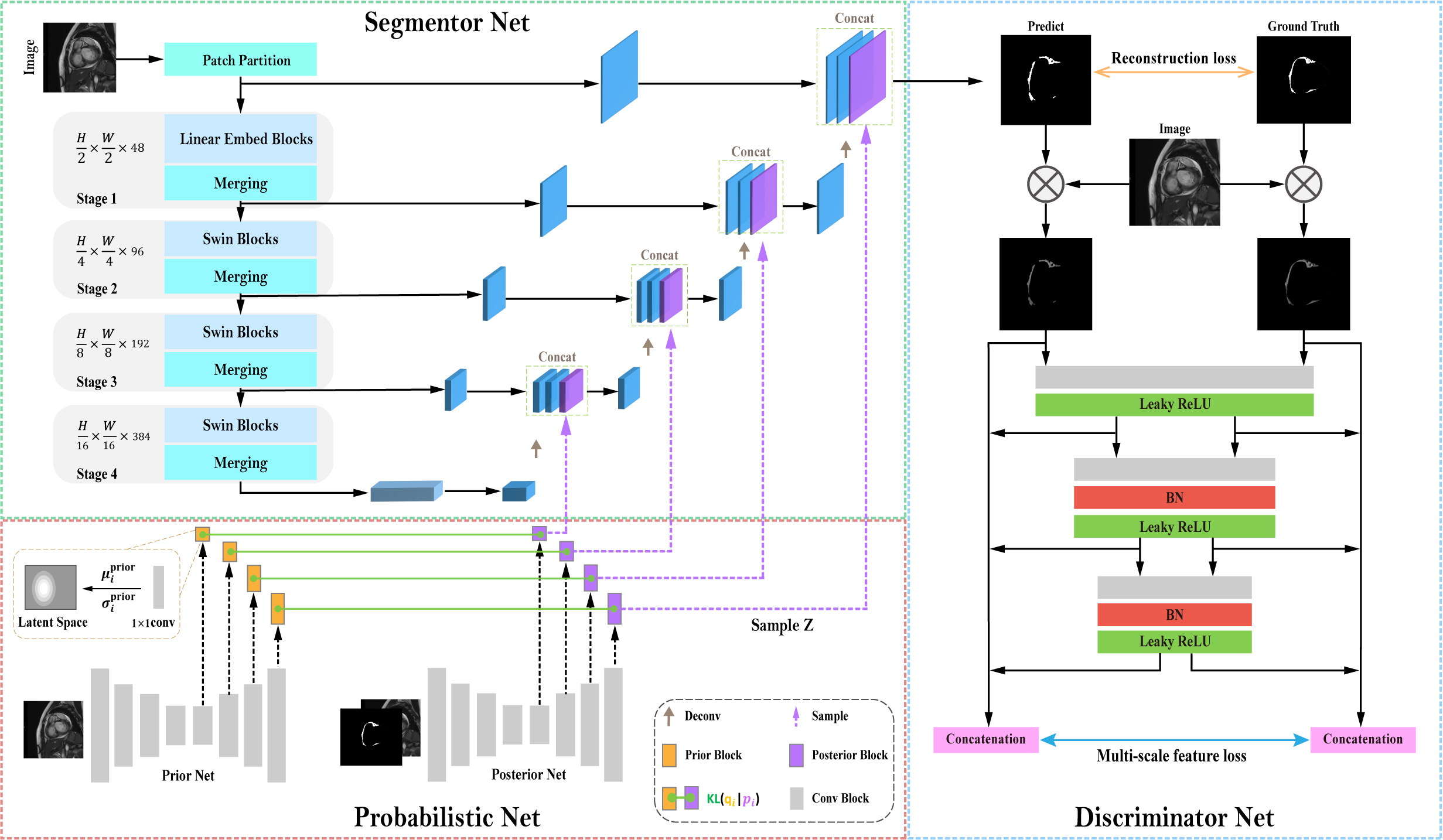}
    {\footnotesize \textbf{Fig2.}}
    {\footnotesize The network is divided into three modules in total: (1) Segmentor Net is used to extract features from the input image and can fuse random feature vectors sampled from (2) Probabilistic Net, and finally get a coarse segmentation. (3)Discriminator Net fuses the coarse segmentation and the true labeling with the input image, respectively, and then computes the loss from these two different inputs.}
    \label{fig:enter-label}
\end{figure}

\subsection*{Materials}
\textbf{Cohort study of EAT Dataset:} The real-word EAT MRI dataset was obtained at the First Affiliated Hospital of South China University by serial image acquisition during end-expiratory breath-holding on a 3.0T scanner (Siemens, Germany). The dataset has been de-identified and was approved by Review Board of Clinical Research 4310 Program of the University of South China (2021KS-XN-12-02) on November 12, 2021. This study adhered to the principles of the Declaration of Helsinki. 

Employing a balanced steady-state free precession (bSSFP) sequence, we acquired 9 to 13 consecutive short-axis cine images of the left ventricle, ranging from the base to the apex. The scanning parameters were as follows: repetition time (TR) / echo time (TE) of 3.30/1.45MS, a field of view of 325×400 mm², flip angle of 45-50°, slice thickness of 8 mm, and a matrix size of 256×166. The raw images were loaded into the commercial software CVI 42 (Circle Cardiovascular Imaging Inc., Calgary, Alberta, Canada). Image post-processing was then performed by an experienced radiologist and two experienced graduate students. Epicardial fat, which is the adipose tissue located between the outer wall of the myocardium and the visceral layer of the pericardium, was manually outlined on left ventricular end-diastolic short-axis images from the base of the left ventricle to the apex. A total of 61 patients' MRI were collected as a training set(about 730 images), and the test set(about 120 images) contained another 10 patients not present in the training set to validate the reliability and effectiveness of our method.

\textbf{ACDC Dataset:} The dataset used in this study is a publicly accessible resource designed for diagnosing cardiac diseases. It encompasses cardiac MRI from 150 patients. Each patient's image sequence covers a complete cardiac cycle, comprising between 28 to 40 images, with a spatial resolution ranging from 1.37 to 1.68 mm²/pixel. Alongside the raw MRI, the ACDC dataset also offers segmentation labels for manual annotation of cardiac structures, encompassing regions of the left ventricle (LV), right ventricle (RV), and interventricular septum (Myo). The training set incorporates data from 100 patients, while the test set includes data from an additional 50 patients. The dataset's richness and diversity render it a pivotal resource for advancing research in the field of cardiac imaging. Moreover, it furnishes a robust experimental foundation for our study. The data can be accessed from https://www.creatis.insa-lyon.fr/Challenge/acdc/.

\subsection*{The SPDNet for EAT Segmentation}
Due to SwinUNETR’s\cite{r23} exceptional performance in medical image segmentation tasks, our method uses it as the Segmentor Net, to achieve more accurate segmentation in MRI-scanned images. Building upon this foundation, we amalgamated the feature information extracted from the Segmentor Net with uncertainty information sampled from the a posteriori feature latent space different scales of the Probabilistic Net.

The approach is shown in Fig. 2, where we combine Probabilistic Net with Segmentor Net. The Probabilistic Net is adept at modeling intricate distributions, whereas the Segmentor Net demonstrates excellent performance in various medical applications. Image and Image + Ground Truth are first modeled by Priori Net and Posteriori Net in Probabilistic Net to construct a priori feature latent space and a posteriori feature latent space, respectively. Next, we sample a random feature vector Z from the a posteriori feature latent space and inject it into the Segmentor Net, which is fused with the Image features extracted by the Segmentor Net to each layer to generate the corresponding segmentation maps and compute the reconstruction loss. Lastly, we employ an adversarial training strategy to train the model through a Discriminator Net with multiscale loss to compensate for the lack of ability to perform direct learning of multiscale space constraints during the end-to-end training process to calibrate the segmentation feature maps. 

\subsubsection*{Segmentor Net}
The backbone network SwinUNETR used an encoder-decoder paradigm and has an overall U-shaped structure, which is derived from the classical model U-Net in the field of medical image segmentation. The Encoder in SwinUNETR is composed of a Patch Partition layer and four stages. The Patch Partition layer divides the image into individual 4$\times$4 sized patches, with each patch treated as a token. These tokens are then processed through Merging and Linear Embed Blocks (comprising a linear embedding layer and Swin Transformer Blocks), denoted as "Stage 1".Swin Transformer Blocks \cite{r24} can the capability to partition the input into non-overlapping windows and perform local self-attention within each window region. The Merging process is employed to amalgamate the patches and augment the dimensionality of the output. In “Stage 1”, there is a reduction in the number of tokens in a single dimension (downsampling by a factor of 2 in resolution), with the output dimension set to 48. Subsequently, the Swin Transformer Blocks and Merging process are reiterated in Stage 2, with the output dimension doubled in comparison to “Stage 1”. This process is repeated twice more in “Stage 3” and “Stage 4”, resulting in output resolutions of \(\frac{H}{8}\times\frac{W}{8}\) and \(\frac{H}{16}\times\frac{W}{16}\), respectively. The features derived from each “Stage” undergo transmission to the CNN-based decode block through skip connections, and “Concat” (the features are spliced together according to the dimensions of the channels) with the features passed from “Deconv”. In the last layer, the features from the Patch Partition are concatenated with the transposed convolution features and features sampled from the Probabilistic Net feature latent space. This concatenation of the features together is based on the dimensions of the channels. Subsequently, a SoftMax is applied to obtain the final segmentation map. Tiny Swin Transformer Blocks (the number of Swin Transformer Blocks for the four stages is [2,2,6,2]) and convolutional layers with the number of channels [48,96,192,384] are used in our approach. This block yields an initial coarse segmentation result, and based on this result, we calculate the reconstruction loss once. The original SwinUNTER loss function comprises the cross-entropy loss function, denoted as \(L_{ce}\left(\hat{y}_{i},y_{i}\right) = \sum^{N}_{i=0}y_{i}\ln\hat{y}_{i}+\left(1-y_{i}\right)\ln\left(1-\hat{y}_{i}\right)\).To capture small lesion features, we introduced the Dice loss, \(L_{dice}\left(\hat{y}_{i},y_{i}\right) = 1-\frac{2\ast \sum^{N}_{i=0}\hat{y}_{i}\ast y_{i}}{\sum^{N}_{i=0}\hat{y}_{i}+\sum^{N}_{i=0}y_{i}}\). These two losses are combined using an adjustable parameter to form a new loss function, denoted as \(L_{rec}\):
\vspace{1\baselineskip}
\begin{equation}
\begin{split}
L_{rec}\left(\hat{y}_{i},y_{i}\right) = -\alpha L_{ce}\left(\hat{y}_{i},y_{i}\right)+\left(1-\alpha\right)L_{dice}\left(\hat{y}_{i},y_{i}\right) \\    
\end{split}    
\end{equation}

In the above equation, where \(i\in\left\{ 0, 1\right\}\) represents the category (0 for background, 1 for epicardial fat), \(\hat{y}_{i}\) denotes the prediction for category \(i\), and \(y_{i}\) denotes the ground truth of class. The parameter \(\alpha\) is artificially set to regulate the ratio between the cross-entropy loss and the Dice loss. In experiments, we find the best performance at $\alpha$ $=$ 0.6.

\subsubsection*{Probabilistic  Net}
Within this module, two U-Net networks with the same structure, namely the a priori net and the posterior net, are incorporated. The primary function of the a priori net is to model the image \(X\). At the decoding stage, multiple low-dimensional latent spaces of different scales, denoted as multiple a priori probability distributions \(p_{i}\left(z_{i}|x_{i}\right)\), are obtained. Each position in these distributions corresponds to a segmentation variable. These are Gaussian distributions with mean \(\mu_{i}^{prior}\) and variance \(\sigma_{i}^{prior}\in\mathbf{\mathbb{R}}^{H_{i} \times W_{i}}\) (where \(i\leq L\), and \( L\)) is the number of layers of the decoder). We compute the \(\mu_{i}^{prior}\) and \(\sigma_{i}^{prior}\) for the feature latent space at each scale by 1$\times$1 convolution. The a priori net comprises a deterministic feature extractor, U-Net, responsible for computing spatial features of a given input image \( X\) at different levels. In a conventional U-Net, the U-Net decoder features at each scale are upsampled and then connected to the U-Net encoder features at the corresponding scale. However, in a priori net, the latent hierarchy at each scale undergoes an additional step: sampling \( z_{i}\in\mathbb{R}^{H_{i} \times W_{i}}\) from the low-dimensional latent space constructed via \(\mu_{i}^{prior}\) and \(\sigma_{i}^{prior}\), which is connected to the features of the input U-Net decoder and then up-sampled in the usual way. By combining the deterministic feature extractor and latent space sampling, we achieve effective utilization of uncertainty at different scales, laying a solid foundation for the accuracy of the segmentation task. The latent variables \( z_{i}\) sampled from the corresponding Gaussian distribution in the latent space and the prior distribution \( P\) are formulated as follows:
\begin{equation}
    \begin{split}
        z_{i}\mathcal{\sim N}\left(\mu_{i}^{prior} \left(x_{i}\right), \sigma_{i}^{prior}\left(x_{i}\right)\right) = p\left(z_{i}|x_{i}\right).
    \end{split}
\end{equation}

\begin{equation}
    \begin{split}
        P\left(z_{0}, \cdots ,z_{i}|x_{0},\cdots ,x_{i}\right) = p\left(z_{0}|x_{0}\right)\cdot \cdot \cdot p\left(z_{i}|x_{i}\right).
    \end{split}
\end{equation}

The posterior net is almost identical in structure to the priori net, differing solely in the number of channels in the input section. It models the variational a posteriori \(Q\left(.|X,Y\right)\) by requiring inputs of both \(X\) and  \(Y\). In our case, \(Y\) is the segmentation label, and \(X\) is the input image. Its role is intended to help maximize the evidence lower bound (ELBO) of the likelihood\textit{ }\( p\left(Y|X\right)\).

In the posterior net, we compute the feature mean \(\mu_{i}^{post}\mathbb{\in R}^{H_{i} \times W_{i}}\) and variance  \(\sigma_{i}^{post}\mathbb{\in R}^{H_{i} \times W_{i}}\) at each scale and model a Gaussian distribution with these means and variances. During the training period, the sampled samples \(z_{i}\) are fed into the Segmentor Net to be fused with the features extracted by the Segmentor Net encoder in the jump connection part. The purpose of this step is to reduce the interference of motion artifacts on the network by using uncertainty constraints to guide the loss function \(L_{rec}\) during the training phase. In the posterior net, the posterior latent variable \(z_{i}\) and the posterior distribution \(P\) are formulated as follows:
\begin{equation}
    \begin{split}
        z_{i}\mathcal{\sim N}\left(\mu_{i}^{post}\left(x_{i},y_{i}\right), \sigma_{i}^{post}\left(x_{i},y_{i}\right)\right) = q\left(z_{i}|x_{i},y_{i}\right)
    \end{split}
\end{equation}

\begin{equation}
    \begin{split}
        Q\left(z_{0}, \cdots, z_{i}|\left(x_{0}, \cdots ,x_{i}\right), \left(y_{0},\cdots ,y_{i}\right)\right) = q\left(z_{0}|x_{0},y_{0}\right)\cdot \cdot \cdot q\left(z_{i}|x_{i},y_{i}\right).
    \end{split}
\end{equation}

In addition, there is a Kullback-Leibler divergence \(D_{KL}\left(Q\vert \vert P\right) = E_{z\sim Q}\left[logQ-logP\right]\),which brings \(P\) and  \(Q\) closer together. We also use a weight parameter \(\beta\) to participate in the computation of the reconstruction loss during the coarse segmentation loss computation (We found that the proposed method works robustly if \(\beta\ = 10\)), and our loss function formula is shown below:

\begin{equation}
    \begin{split}
        L_{ELBO} = {E_{z\sim Q}}\left[L_{rec}\left(S\left(X,z\right),Y\right)\right]+\beta \cdot \sum_{i = 0}^{L}{E_{z_{i}\sim Q}D_{KL}}\left(q_{i}\left(z_{i}|x_{i},y_{i}\right)\parallel p_{i}\left(z_{i}|x_{i}\right)\right).
    \end{split}
\end{equation}

\subsubsection*{Discriminator   Net}
Discriminator Net contains multilayer convolutional layers and leaky activation functions, while batch normalization is introduced in the last two convolutional layers to normalize the batch data. 

It has two different inputs: the coarse segmented image \(S\left(X,z\right)\) fused with the input image \(X\) and the segmented label \(Y\) fused with the input image \(X\) denoted as \(D\left(X\ast S\left(X,z\right)\right)\) and \(D\left(X\ast Y\right)\). The Discriminator Net employs a multilayer structure to extract hierarchical features and compute a multiscale loss \(L_{1}\). This loss makes full use of the different scale of features, allowing the model i.e. to learn global and local features more efficiently and to calibrate the segmented feature map further. Training of the Discriminator Net occurs in an alternating manner with the Segmentor Net through backpropagation of the multiscale loss \(L_{1}\). During training, the parameters of the Segmentor Net are initially fixed to update the parameters of the Discriminator Net. Subsequently, the parameters of the Discriminator Net are fixed, and the parameters of the Segmentor Net are updated by computing the gradient using the losses propagated from the Discriminator Net. This training approach is akin to a minimax game: the objective of the Segmentor Net is to minimize the multi-scale feature loss during training, whereas the Discriminator Net endeavors to maximize the multi-scale feature loss.

Through continual training and parameter updates, the Discriminator Net and Segmentor Net will become more and more powerful and gradually reach a balance. In Discriminator Net we use Mean Absolute Error (MAE) \(L_{mse}\) for multi-scale feature loss, so our Discriminator Net loss is as follows:

\begin{equation}
    \begin{split}
        L_{mse}\left(D\left(X\ast Y\right), D\left(X\ast S\left(X,z\right)\right)\right) = \frac{1}{C}\sum_{i = 1}^{C}\vert\vert\left(d_{i}\left(X\ast Y\right), d_{i}\left(X\ast S\left(X,z\right)\right)\right)\vert\vert_{1}.
    \end{split}
\end{equation}

In the above equation, \(C\) is the total number of layers of the Discriminator Net, and \(d_{i}\) represents the features extracted from the \(i\)th convolutional layer. 

In the following expression, \(\theta_{S}\) and \(\theta_{D}\) represent the parameters of the generator and discriminator, respectively, and K is the batch size of the input image. To better the optimization of our objective function, we integrate the ELBO loss with the discriminator training loss. Thus, the total loss of our model is expressed as follows:

\begin{equation}
    \begin{split}
        \min_{\theta_{S}}\max_{\theta_{D}}L_{total}\left(\theta_{S},\theta_{D}\right) = \frac{1}{K}\sum_{i=1}^{K}L_{mse}\left(D\left(X\ast Y\right),D\left(X\ast S\left(X,z\right)\right)\right)+L_{ELBO}.
    \end{split}
\end{equation}

\section*{Experiments and Results}
In this study, we conducted a comprehensive performance evaluation of the model using the ACDC MICCAI 2017 cardiac dataset and the real-world epicardial fat dataset. We deliberately selected six contemporary and leading medical image segmentation methods: SegAN \cite{r21}, FCT\cite{r25}, nnFormer\cite{r26}, SwinUnet\cite{r27}, TransUNet\cite{r28}, and U-Net\cite{r7}for comparative experiments.
To assess the segmentation performance, three metrics were employed: Dice score, Jaccard (Jaccard similarity coefficient), and Harsdorff distance (HD). The model underwent training for 100 epochs using Adam's optimization function, with a batch size set to 8 and an initial learning rate of 0.0001. Throughout the training process, we utilized an NVIDIA 3090 graphics card and enhanced the model's performance with data augmentation techniques, including random rotations, flipping samples on either the x- or y-axis, and introducing random angular skews. Such experimental design ensures the reliability and rigor of our comparison tests and provides a solid foundation for the robustness of the method in practical applications.

\begin{figure}
    \centering
    \includegraphics[width=1\linewidth]{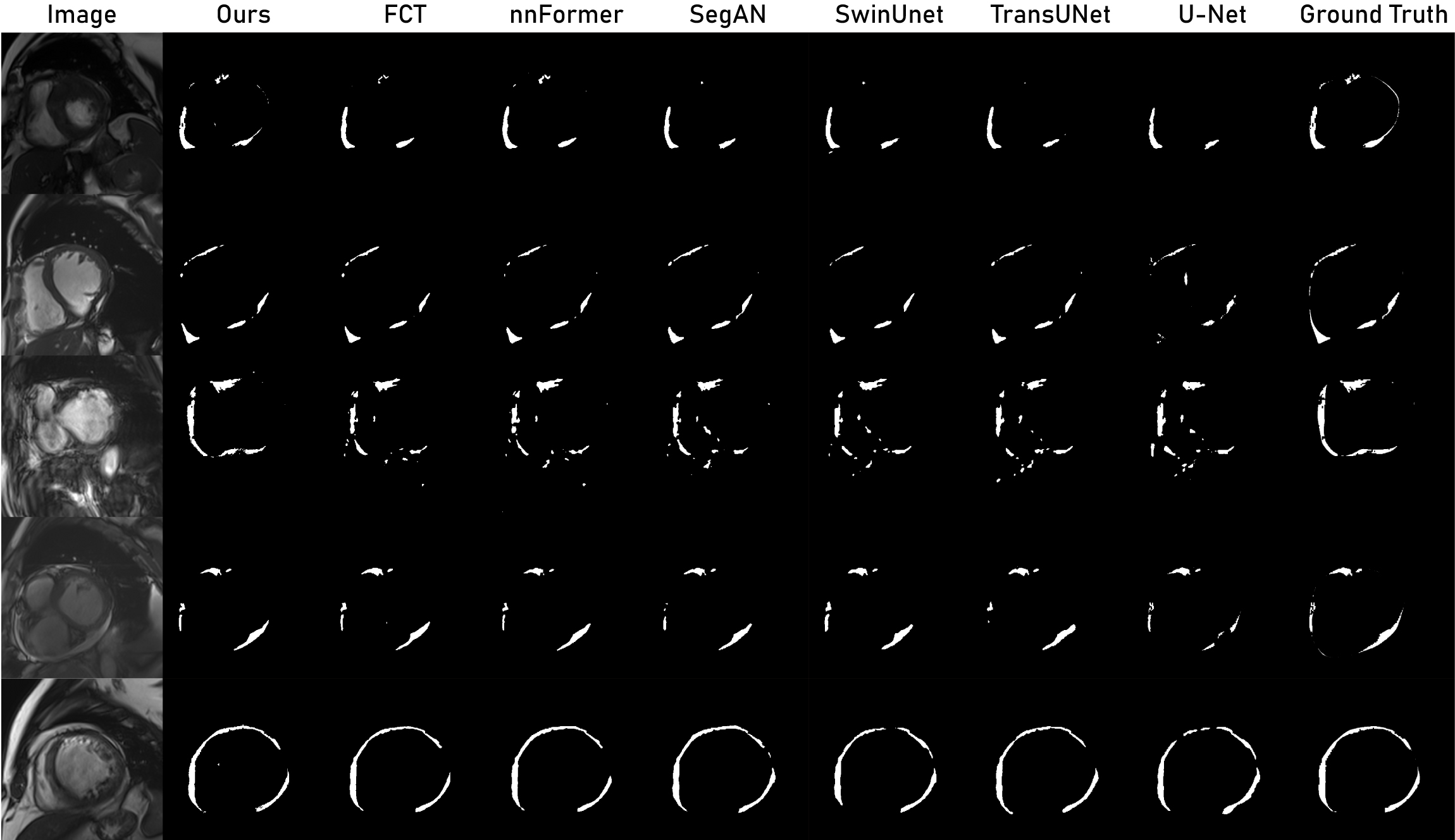}
    {\footnotesize \textbf{Fig3.}}
    {\footnotesize The visualized of the epicardial fat segmentation results, with each row in the first column being a different MRI slice of the patient's heart. We compared several other different methods and compared the model predictions with the Ground Truth, with the last column being GT.}
    \label{fig:enter-label}
\end{figure}

\begin{figure}
    \centering
    \includegraphics[width=1\linewidth]{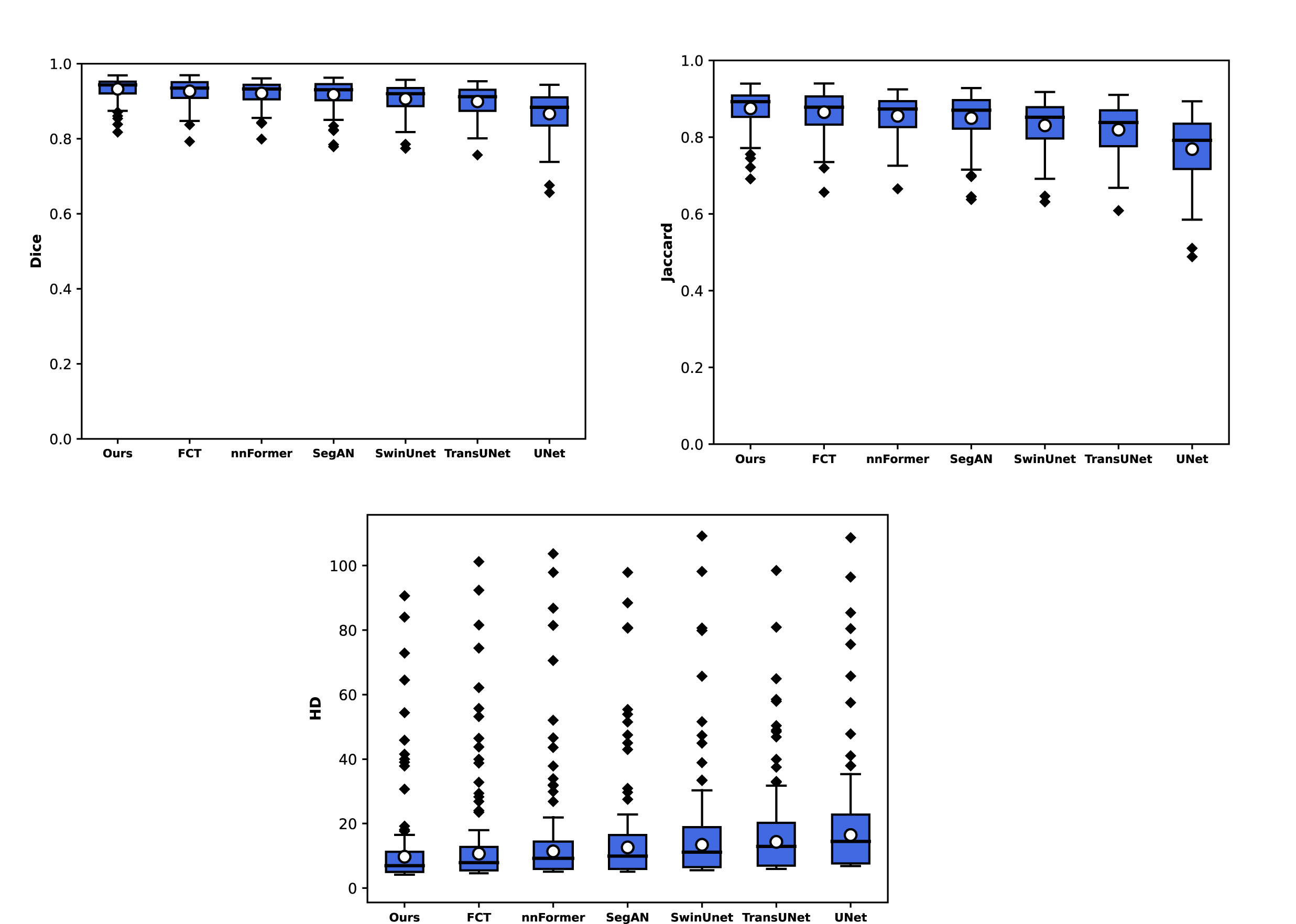}
    {\footnotesize \textbf{Fig4.}}
    {\footnotesize The segmentation performance of our proposed method as well as other methods is compared in the test set based on Dice score, Jaccard (Jaccard similarity coefficient) and Hausdorff Distance (HD) scores and visualized in the form of box plots. The x and y axes of each plot indicate the name of the model and the evaluation scores for each metric, respectively.}
    \label{fig:enter-label}
\end{figure}

\subsection*{Comparative experiments with other segmentation models}
According to Table 1, we conducted a comprehensive comparison of our model with six other models on the private dataset. The results demonstrate that our model achieves optimal scores on all three evaluation metrics. Specifically, the Dice score reaches 0.933, the Jaccard coefficient is 0.875, and the Harsdorff distance is 9.696mm. Additionally, we visually represented the distribution of the evaluation metric scores through box-and-line plots in Fig. 4. The results clearly indicate that, compared to the other methods, the model significantly reduces the number of outliers predicted for the three metrics, further confirming the stability and superiority of our method. These results fully showcase the excellent performance of our model in medical image segmentation tasks.

Observing Fig. 3, we note a significant segmentation deficiency in other methods, particularly in cases of motion blur. Additionally, in images where epicardial fat and pericardial effusion coexist, other methods encounter difficulty in differentiation. This underscores the model's effective mitigation of the effects of motion blur and its enhanced ability to accurately distinguish epicardial fat from pericardial effusion.

To further validate the segmentation performance, we conducted experiments on the public dataset ACDC MICCAI 2017, with the results presented in Table 2. On this dataset, the model also achieves the best performance across various metrics. These experiments demonstrate that the method is not only feasible in the cardiac binary classification task but also effective in the cardiac multiclassification task. We visualized the segmentation results by differentiating the categories through different colors, as shown in Fig. 5. And we represented the distribution of the evaluation metric scores in Fig. 6. This series of results clearly demonstrate the excellent performance of our model, providing a reliable solution for cardiac image segmentation.

\begin{figure}
    \centering
    \includegraphics[width=1\linewidth]{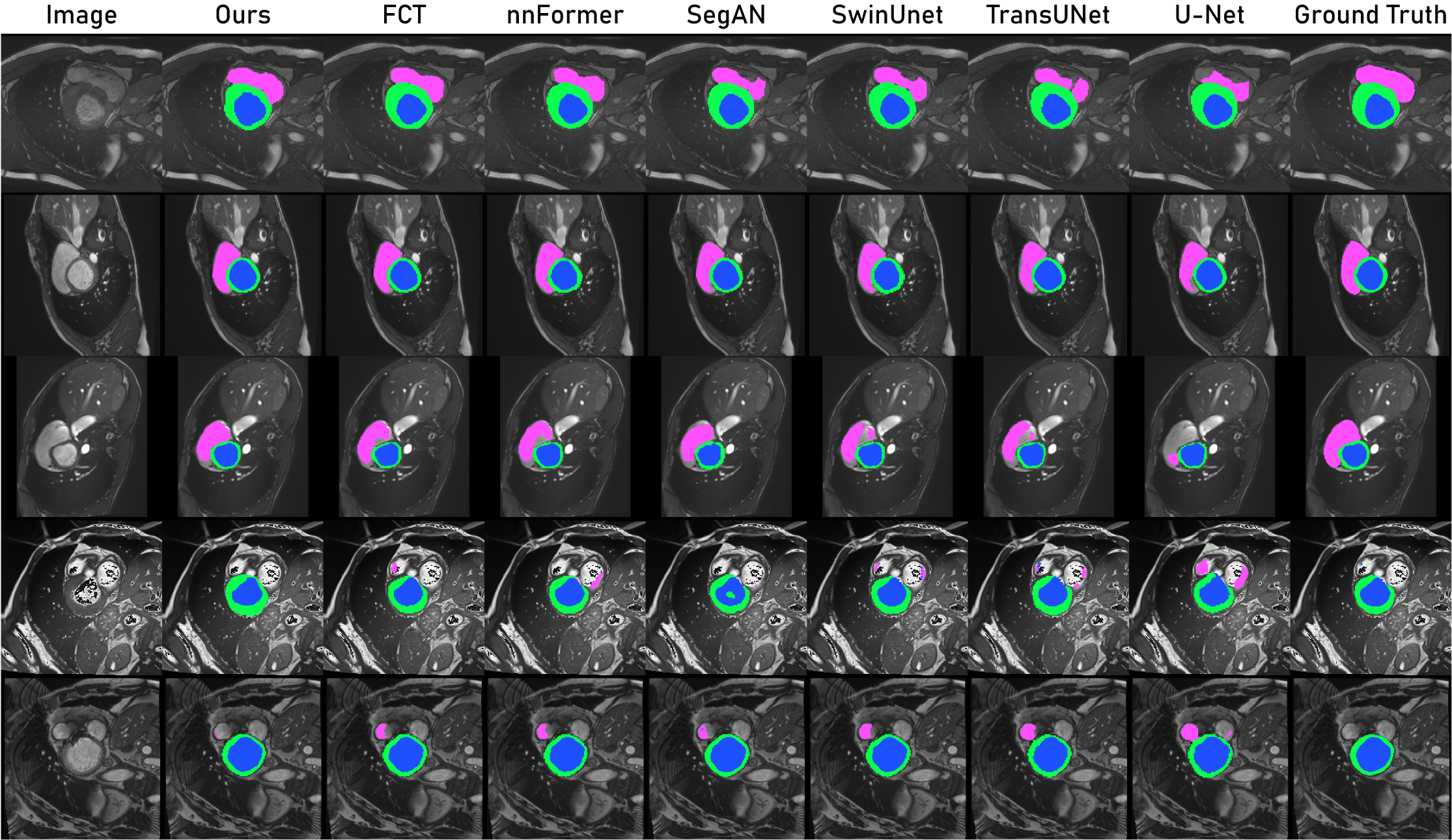}
    {\footnotesize \textbf{Fig5.}}
    {\footnotesize Visualization of segmentation results in the ACDC dataset compared to other methods, with the left ventricle in blue, the left ventricular myocardium in green, and the right ventricle in pink.}
    \label{fig:enter-label}
\end{figure}

\begin{figure}
    \centering
    \includegraphics[width=1\linewidth]{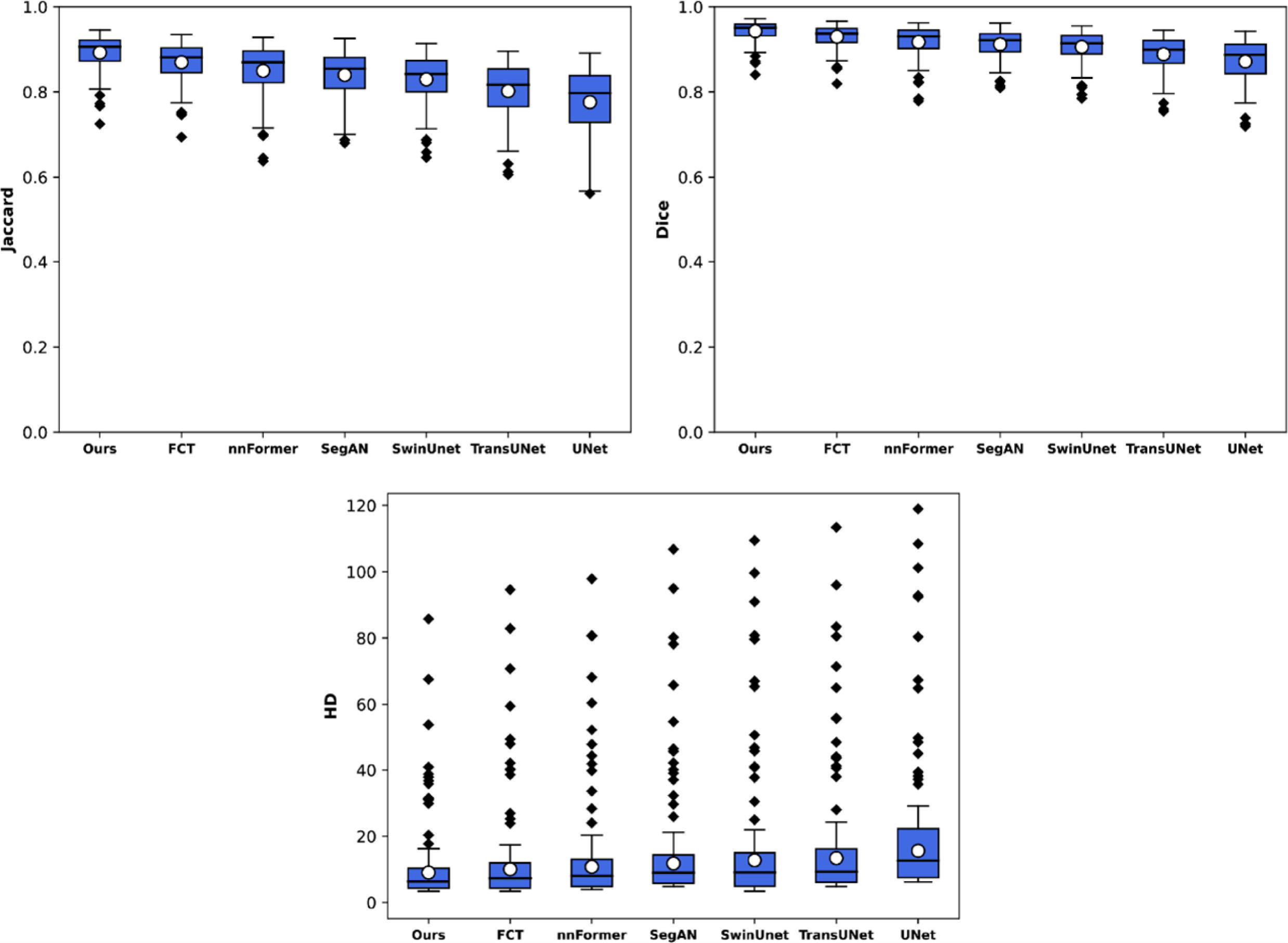}
    {\footnotesize \textbf{Fig6.}}
    {\footnotesize Boxplot of each indicator in the ACDC test set.}
    \label{fig:enter-label}
\end{figure}

\vspace{1\baselineskip}
\begin{table}[H]
    \begin{adjustbox}{max width=\textwidth}
        \begin{tabular}{p{4.0cm}p{3.55cm}p{3.55cm}p{3.55cm}}
            \hline
            \multicolumn{1}{p{4.0cm}}{Methods} &
            \multicolumn{1}{|p{3.55cm}}{Dice score} &
            \multicolumn{1}{|p{3.55cm}}{Jaccard} &
            \multicolumn{1}{|p{3.55cm}}{HD(mm)} \\
            \hline
            \multicolumn{1}{p{4.0cm}}{U-Net\cite{r7}} &
            \multicolumn{1}{|p{3.55cm}}{0.866} &
            \multicolumn{1}{|p{3.55cm}}{0.771} &
            \multicolumn{1}{|p{3.55cm}}{13.721} \\

            \multicolumn{1}{p{4.0cm}}{TransUNet\cite{r28}} &
            \multicolumn{1}{|p{3.55cm}}{0.899} &
            \multicolumn{1}{|p{3.55cm}}{0.822} &
            \multicolumn{1}{|p{3.55cm}}{12.617} \\

            \multicolumn{1}{p{4.0cm}}{SwinUNet\cite{r27}} &
            \multicolumn{1}{|p{3.55cm}}{0.906} &
            \multicolumn{1}{|p{3.55cm}}{0.831} &
            \multicolumn{1}{|p{3.55cm}}{12.503} \\

            \multicolumn{1}{p{4.0cm}}{SegAN\cite{r21}} &
            \multicolumn{1}{|p{3.55cm}}{0.917} &
            \multicolumn{1}{|p{3.55cm}}{0.849} &
            \multicolumn{1}{|p{3.55cm}}{11.368} \\

            \multicolumn{1}{p{4.0cm}}{nnFormer\cite{r26}} &
            \multicolumn{1}{|p{3.55cm}}{0.921} &
            \multicolumn{1}{|p{3.55cm}}{0.855} &
            \multicolumn{1}{|p{3.55cm}}{10.683} \\
            
            \multicolumn{1}{p{4.0cm}}{FCT\cite{r25}} &
            \multicolumn{1}{|p{3.55cm}}{0.927} &
            \multicolumn{1}{|p{3.55cm}}{0.865} &
            \multicolumn{1}{|p{3.55cm}}{10.356} \\

            \multicolumn{1}{p{4.0cm}}{\textbf{SPDNet(ours)}} &
            \multicolumn{1}{|p{3.55cm}}{\textbf{0.933}} &
            \multicolumn{1}{|p{3.55cm}}{\textbf{0.875}} &
            \multicolumn{1}{|p{3.55cm}}{\textbf{9.696}} \\
            \hline
        \end{tabular}
    \end{adjustbox}
\end{table}
\noindent
{\footnotesize \textbf{Table 1}}
{\footnotesize Comparative experiment results with the segmentation methods in the table were evaluated on the epicardial fat dataset. The evaluation metrics used are Dice score, Jaccard (Jaccard similarity coefficient) and Hausdorff Distance (HD), with black bolding representing the best performance.}

\vspace{1\baselineskip}
\begin{table}[H]
    \begin{adjustbox}{max width=\textwidth}
        \begin{tabular}{p{4.0cm}p{3.55cm}p{3.55cm}p{3.55cm}}
            \hline
            \multicolumn{1}{p{4.0cm}}{Methods} &
            \multicolumn{1}{|p{3.55cm}}{Dice score} &
            \multicolumn{1}{|p{3.55cm}}{Jaccard} &
            \multicolumn{1}{|p{3.55cm}}{HD(mm)} \\
            \hline
            \multicolumn{1}{p{4.0cm}}{U-Net\cite{r7}} &
            \multicolumn{1}{|p{3.55cm}}{0.876} &
            \multicolumn{1}{|p{3.55cm}}{0.786} &
            \multicolumn{1}{|p{3.55cm}}{14.174} \\

            \multicolumn{1}{p{4.0cm}}{TransUNet\cite{r28}} &
            \multicolumn{1}{|p{3.55cm}}{0.897} &
            \multicolumn{1}{|p{3.55cm}}{0.824} &
            \multicolumn{1}{|p{3.55cm}}{12.429} \\

            \multicolumn{1}{p{4.0cm}}{SwinUNet\cite{r27}} &
            \multicolumn{1}{|p{3.55cm}}{0.901} &
            \multicolumn{1}{|p{3.55cm}}{0.831} &
            \multicolumn{1}{|p{3.55cm}}{11.792} \\

            \multicolumn{1}{p{4.0cm}}{SegAN\cite{r21}} &
            \multicolumn{1}{|p{3.55cm}}{0.912} &
            \multicolumn{1}{|p{3.55cm}}{0.854} &
            \multicolumn{1}{|p{3.55cm}}{10.968} \\

            \multicolumn{1}{p{4.0cm}}{nnFormer\cite{r26}} &
            \multicolumn{1}{|p{3.55cm}}{0.921} &
            \multicolumn{1}{|p{3.55cm}}{0.864} &
            \multicolumn{1}{|p{3.55cm}}{10.053} \\
            
            \multicolumn{1}{p{4.0cm}}{FCT\cite{r25}} &
            \multicolumn{1}{|p{3.55cm}}{0.931} &
            \multicolumn{1}{|p{3.55cm}}{0.873} &
            \multicolumn{1}{|p{3.55cm}}{9.818} \\

            \multicolumn{1}{p{4.0cm}}{\textbf{SPDNet(ours)}} &
            \multicolumn{1}{|p{3.55cm}}{\textbf{0.939}} &
            \multicolumn{1}{|p{3.55cm}}{\textbf{0.881}} &
            \multicolumn{1}{|p{3.55cm}}{\textbf{9.049}} \\
            \hline
        \end{tabular}
    \end{adjustbox}
\end{table}
\noindent
{\footnotesize \textbf{Table2}}
{\footnotesize Comparative experiment results with other methods on the ACDC public dataset.}
\vspace{1\baselineskip}

\begin{figure}
    \centering
    \includegraphics[width=1\linewidth]{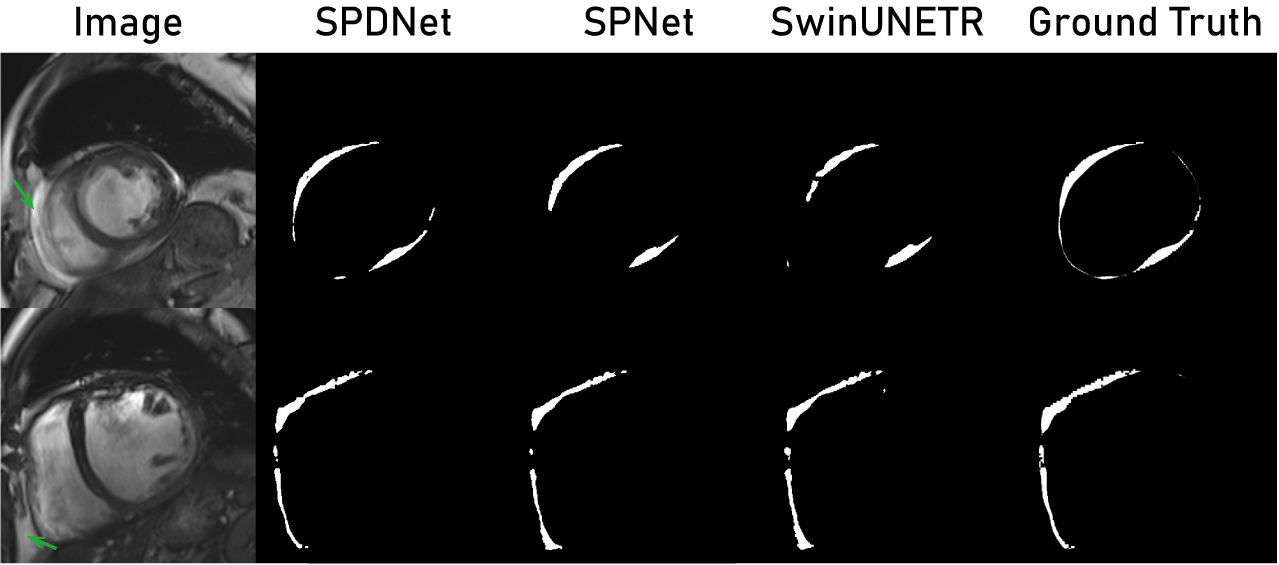}
    {\footnotesize \textbf{Fig7.}}
    {\footnotesize Results from ablation experiments were visually presented in case sections containing either motion artifacts or pericardial effusions. The first row of case sections illustrates instances with motion artifacts, while the second row showcases sections where pericardial effusion is present in proximity to epicardial adipose tissue. Relevant areas have been highlighted with green arrows for clarity.}
    \label{fig:enter-label}
\end{figure}

\subsection*{Ablation experiments}
The performance improvement of our algorithm lies in the introduction of two key nets: the Probabilistic Net and the Discriminator Net. To assess the effectiveness of these nets, we conducted exhaustive ablation experiments, except for adding or subtracting net, we did not modify other aspects (in Fig. 7).

The results of the experiments, as presented in Table 3, indicate a significant enhancement in the model's performance with the expansion of each net. Introducing the Probabilistic Net results in a 0.03 improvement in the Dice coefficient score, a 0.03 improvement in the Jaccard index, and a 0.39 mm improvement in the Harsdorff distance index. This experiment substantiates the efficacy of our enhancements. Similarly, the addition of the Discriminator Net further enhances the performance across each metric. The improvements observed in these quantitative metrics underscore the effectiveness of our enhancement measures. These results robustly support the superior performance of our model and the validity of our methods.

\vspace{2\baselineskip}
\begin{table}[H]
    \begin{adjustbox}{max width=\textwidth}
        \begin{tabular}{p{4.05cm}p{3.37cm}p{2.82}p{2.16cm}p{2.16cm}}
            \hline
            \multicolumn{1}{p{4.05cm}}{Probabilistic Net} &
            \multicolumn{1}{|p{3.37cm}}{Segmentor Net} &
            \multicolumn{1}{|p{2.82cm}}{Dice score} &
            \multicolumn{1}{|p{2.16cm}}{Jaccard } &
            \multicolumn{1}{|p{2.16cm}}{HD(mm)} \\
            \hline
            
            \multicolumn{1}{p{4.05cm}}{} &
            \multicolumn{1}{|p{3.37cm}}{} &
            \multicolumn{1}{|p{2.82cm}}{0.921} &
            \multicolumn{1}{|p{2.16cm}}{0.857} &
            \multicolumn{1}{|p{2.16cm}}{10.679} \\
            
            \multicolumn{1}{p{4.05cm}}{$\surd$} &
            \multicolumn{1}{|p{3.37cm}}{} &
            \multicolumn{1}{|p{2.82cm}}{0.925} &
            \multicolumn{1}{|p{2.16cm}}{0.863} &
            \multicolumn{1}{|p{2.16cm}}{10.287} \\

            \multicolumn{1}{p{4.05cm}}{$\surd$} &
            \multicolumn{1}{|p{3.37cm}}{$\surd$} &
            \multicolumn{1}{|p{2.82cm}}{0.933} &
            \multicolumn{1}{|p{2.16cm}}{0.875} &
            \multicolumn{1}{|p{2.16cm}}{9.696} \\
            \hline
        \end{tabular}
    \end{adjustbox}
\end{table}
\noindent
{\footnotesize \textbf{Table3}}
{\footnotesize Performance of ablation experiments with different modules.}
\vspace{1\baselineskip}

\subsection*{Limitations and Discussion}
This study focuses on solving the problem of motion artifacts appearing in MRI and quantifying them as uncertainty. However, our approach is limited by the specific problem and application scenarios and thus has some limitations in considering uncertainty. Furthermore, we used a Bayesian-based approach to construct a Probabilistic Net, which learns the distribution of network weights with uncertainty. While this provides us with more comprehensive information, it also comes with a larger computational cost.

In future endeavors, we aim to enhance our methods to cater to a broader spectrum of application scenarios and streamline computational efficiency by optimizing our algorithms. This involves investigating more efficient approaches to quantify uncertainty and ensuring scalability and practicality. Additionally, we are committed to designing a series of new experiments to comprehensively evaluate the effectiveness of uncertainty and validate the performance of our methods across diverse contexts.

\section*{Conclusions}
In this study, epicardial fat segmentation faced two major challenges: the difficulty in distinguishing the presence of motion artifacts and EAT from pericardial effusion. In addition, manual labeling of epicardial fat is a time-consuming and laborious task. Therefore, our study aims to propose an efficient automated segmentation model to assist clinicians in risk assessment and thus improve their productivity. Our approach optimizes the Segmentor Net by transforming the blurring of epicardial adipose tissue (EAT) edges induced by motion artifacts into a regularization constraint via the Probabilistic Net. This optimization enhances segmentation performance through the utilization of uncertainty representation. Subsequently, an adversarial training strategy is employed to learn the predictive segmentation maps and ground truth multiscale feature differences by constructing a Discriminator Net with multiscale loss to calibrate the predictive segmentation maps and better differentiate pericardial effusion from epicardial fat.

The proposed model was experimentally validated on the public cardiac dataset ACDC and the real-world epicardial fat dataset. The experimental results show that the proposed model has achieved significant performance improvement, which provides a reliable auxiliary tool for clinicians to accurately segment epicardial fat and provides strong support for risk assessment and diagnosis.

\section*{Acknowledgement}
This work was supported by National Natural Science Foundation of China [62171166,62272212]; Research Foundation of the First Affiliated Hospital of University of South China for Advanced Talents [20210002-1005 USCAT-2021-01]; Natural Science Foundation of Hunan Province [2023JJ30536,2023JJ30547]; Scientific Research Fund Project of Hunan Provincial Health Commission [20201920]; Special Funds for the Construction of Innovative Provinces in Hunan [2020SK4008]; Clinical Research 4310 Program of the University of South China [20224310NHYCG05,20214310NHYCG03].

\bibliographystyle{unsrt}
\bibliography{ref}
\end{document}